\documentclass[3p,times,procedia]{elsarticle}

\usepackage{ecrc}
\usepackage{graphicx,amsmath,amssymb}

\volume{00}

\firstpage{1}

\runauth{}

\CopyrightLine{2015}{Published by Elsevier Ltd.}

\usepackage[figuresright]{rotating}

\begin{document}

\begin{frontmatter}


\title{Electronic Scattering Effects in Europium-Based Iron Pnictides}

\dochead{}

 \author[label1]{S. Zapf\corref{cor1}}
 \cortext[cor1]{Corresponding author.}
\ead{sina.zapf@pi1.physik.uni-stuttgart.de}
  \author[label1]{D. Neubauer}
  \author[label2]{K. W. Post}
 \author[label1]{A. Kadau}
  \author[label1]{J. Merz}
   \author[label1]{C. Clauss}
    \author[label1]{A.~L\"ohle}
     \author[label3,label4]{H. S. Jeevan}
      \author[label3,label5]{P. Gegenwart}
       \author[label2]{D. N. Basov}
        \author[label1]{M. Dressel}
\address[label1]{1.~Physikalisches Institut, Universit\"at Stuttgart, Pfaffenwaldring 57, 70550 Stuttgart, Germany}
\address[label2]{Department of Physics, University of California, San Diego, La Jolla, California 92093, USA}
\address[label3]{I.~Physikalisches Institut, Universit\"at G\"ottingen, Friedrich-Hund-Platz 1, 37077 G\"ottingen, Germany}
\address[label4]{Department of Physics, PESITM, Sagar Road, 577204 Shimoga, India}
\address[label5]{Experimentalphysik~VI,  Universit\"at Augsburg, Universit\"atsstra{\ss}e 1, 86135 Augsburg, Germany}

\begin{abstract}
In a comprehensive study, we investigate the electronic scattering effects in EuFe$_{2}$(As$_{1-x}$P$_{x}$)$_{2}$ by using Fourier-transform infrared spectroscopy. In spite of the fact that Eu$^{2+}$ local moments order around $T_\text{Eu} \sim 20$\,K, the overall optical response is strikingly similar to the one of the well-known Ba-122 pnictides.
The main difference lies within the suppression of the lower spin-density-wave gap feature. By analysing our spectra with a multi-component model, we find that the high-energy feature around 0.7\,eV -- often associated with Hund's rule coupling -- is highly sensitive to the spin-density-wave ordering; this further confirms its direct relationship to the dynamics of itinerant carriers. The same model is also used to investigate the in-plane anisotropy of magnetically detwinned EuFe$_{2}$As$_{2}$ in the antiferromagnetically ordered state, yielding a higher Drude weight and lower scattering rate along the crystallographic $a$-axis. Finally, we analyse the development of the room temperature spectra with isovalent phosphor substitution and highlight changes in the scattering rate of hole-like carriers induced by a Lifshitz transition.
\end{abstract}


\end{frontmatter}


\section{Introduction}
Iron pnictides are fascinating multiband materials that offer various options how to suppress the antiferromagnetic ground state of the parent compounds in order to introduce superconductivity~\cite{Johnston,materials}. Commonly, one distinguishes between charge-carrier doping, isovalent substitution and mechanical pressure as tuning parameters through the corresponding phase diagrams. However, several experiments indicate that disorder also plays a crucial role; for instance, signs of a quantum critical point hidden beneath the superconducting dome were only found in the case of BaFe$_2$(As$_{1-x}$P$_x$)$_2$, a material which is usually considered as very clean due to the isovalent substitution outside the conducting Fe layers~\cite{Shibauchi_QCP_1, Shibauchi_QCP_2}. With this in mind, putting scattering and disorder effects in the foreground offers a different perspective on the relevant mechanisms responsible for the variation of electronic properties in the phase diagrams of iron-based superconductors~\cite{disorder}.

Infrared spectroscopy has been established as the ideal tool to investigate electronic scattering effects in correlated materials~\cite{Basov_review}. In the case of iron-based superconductors, up to now most of the optical studies were performed on $X$Fe$_2$As$_2$ ``122'' iron pnictides  with $X$ = Ba, because rather large high-quality single crystals were available. However, the choice of $X$ influences the electronic properties remarkably. This is revealed, for example, by the observation that the spin-density-wave transition temperatures $T_\text{SDW}$ do not follow the ionic radii of the substituents~\cite{radii}.

Eu-based compounds are peculiar members of the 122 iron pnictides which display an additional local magnetic order around $T_\text{Eu} \sim 20$\,K, leading to  a unique interplay of magnetic, superconducting and structural effects~\cite{P_2011_Jeevan, P_2011_Nowik}. This interplay manifests itself for example in the lifting of the nodes of the superconducting gap in EuFe$_{2}$(As$_{0.82}$P$_{0.18}$)$_{2}$, probably caused by magnetic  spin scattering on Eu$^{2+}$ moments~\cite{P_2011_Wu}. Moreover, a persistent magnetic detwinning can be induced by small magnetic fields, meaning that once an external magnetic field was applied at temperatures below the structural (nematic) and spin-density-wave transition at $T_\text{s,SDW} = 190$\,K, the crystal stays (partially) detwinned even when the field is switched off~\cite{2014_Zapf}. This allows to investigate the intrinsic electronic in-plane anisotropy of iron pnictides without any mechanical device.

These extraordinary properties motivated us to study the electrodynamic in-plane response of a series of EuFe$_{2}$\-(As$_{1-x}$P$_{x}$)$_{2}$ single crystals, using Fourier-transform infrared (FTIR) spectroscopy. We have chosen the compositions $x = 0$, 0.12, 0.165, 0.26 as representative, as the former two display a spin-density-wave ground state, while the latter two mark the boarders of the rather narrow superconducting dome that appears in europium-based iron pnictides~\cite{P_2011_Jeevan, P_2011_Nowik, P_2013_Zapf}. Here we review and discuss unexpected but very interesting anomalies concerning the electronic scattering in the normal and spin-density-wave state. In the first part of the manuscript, we compare the phase diagrams of EuFe$_{2}$As$_2$ under pressure and isovalently substituted EuFe$_{2}$(As$_{1-x}$P$_{x}$)$_{2}$, leading to the conclusion that crystallographic changes are not the sole key for understanding the phase diagram of these compounds. In the following, we describe how Eu$^{2+}$ spin scattering influences the temperature dependent optical properties of  EuFe$_{2}$As$_2$. Furthermore, we analyse our data with a multi-component model. This allows us to extract quantitative information, how a high-energy feature that is usually ascribed to correlation effects is influenced by the spin-density-wave transition. The same model is also used to analyse the in-plane anisotropy of a detwinned crystal at 30\,K. Finally, we investigate the development of the room-temperature spectra with isovalent P substitution and highlight changes in the scattering rate of incoherent carriers induced by a Lifshitz transition, $i.e.$ a topological transition of the Fermi surface without change in symmetry.

\section{Materials and Methods}
\begin{figure} [!http]
 \centering
  \includegraphics[width=1\columnwidth]{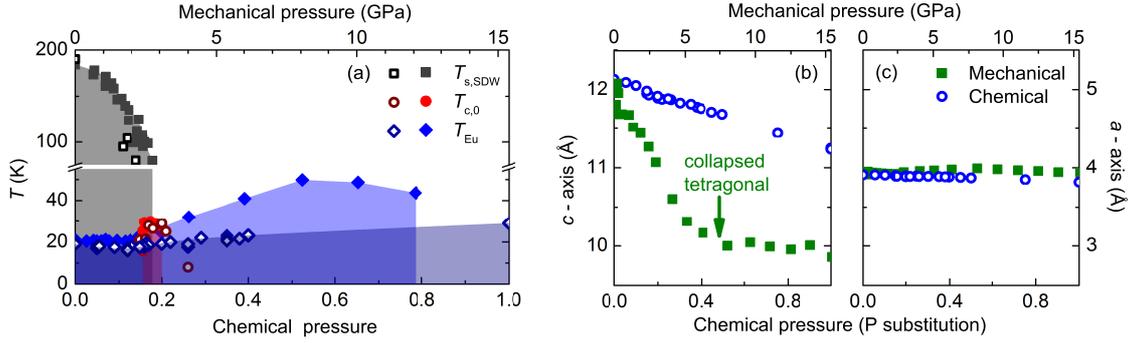}
  \caption{(a)~Unified phase diagrams of EuFe$_2$As$_2$ single crystals under chemical (P substitution, open dark symbols, Ref.~\cite{P_2011_Jeevan, P_2011_Wu, P_2011_Thiru, P_2011_Zapf, PK_2012_Maiwald, othersP_2012_Krug, pressureP_2012_Tokiwa, P_2013_Zapf, P_2013_Jiang, P_2014_Nandi}) and mechanical pressure (closed bright symbols, Ref.~\cite{pressure_2011_Matsubayashi, pressure_2011_Kurita3}) mapped on top of each other, using the relation: P substitution $x = 1 \mathop{\hat{=}}  15.4$\,GPa. While the electronic properties are quite similar, $T_\text{Eu}$ (blue diamonds) develops differently. However, in both cases $T_\text{Eu}$ starts to increase stronger close to the pressure, where $T_\text{s,SDW}$ (black squares) gets suppressed and superconductivity appears ($T_\text{c,0}$: red circles). (b)~Crystallographic $c$-axis (left) and $a$-axis (right) of EuFe$_2$As$_2$ under chemical (P substitution, blue open points, Ref.~\cite{P_2011_Jeevan, P_2011_Cao}) and mechanical pressure (green squares, Ref.~\cite{pressure_2010_Uhoya}). Using the equivalent: P content $x = 1 \mathop{\hat{=}}  15.4$\,GPa that maps the electronic properties on top of each other, reveals that mechanical pressure contracts the $c$-axis much stronger than P substitution.}
  \label{pressureP}
\end{figure}

In the case of iron pnictides, there is a general interest in understanding how chemical substitution and pressure lead to the emergence of superconductivity. Besides the crystallographic structure and doping, disorder was proposed to be an alternative  parameter to tune through their phase diagram~\cite{disorder}. Indeed, when we compare the phase diagram of EuFe$_2$(As$_{1-x}$P$_x$)$_2$ with the one of EuFe$_2$As$_2$ under pressure, it becomes obvious that isovalent P substitution does not only act as mechanical pressure: using the equivalent P content $x = 1 \mathop{\hat{=}}  15.4$\,GPa that maps most efficiently the spin-density-wave and superconducting transition temperatures on top of each other\footnote{This factor is close to the one (12.2\,GPa) determined by a calculation based on the lattice constants and the bulk modulus~\cite{pressureP_2012_Tokiwa}. The present factor guarantees that the here displayed data overlap best over a broad pressure range. The 20\% discrepancy can be explained by a strong dependence of the electronic properties on the hydrostaticity of mechanical pressure which is well-known for Eu-based iron pnictides~\cite{pressure_2011_Matsubayashi}.}, reveals that mechanical pressure contracts the $c$-axis much stronger than P substitution (see Fig.~\ref{pressureP}). Furthermore, also the Eu$^{2+}$ ordering temperature develops differently throughout the phase diagram. Since the unequal evolution of the $c$-parameter should modify the RKKY-exchange, we associate the differences of the Eu$^{2+}$-ordering temperature in the phase diagram with a change of magnetic exchange induced by the modified lattice parameters

The temperature-dependent in-plane optical reflectivity of EuFe$_2$As$_2$ was measured over a wide frequency range from 40 to 10\,000\,cm$^{-1}$, using three  infrared Fourier transform spectrometers (Bruker IFS 66 v/s, IFS 113v and Vertex 80v). As a reference, either gold was deposited on the sample and the measurements were repeated at each temperature (40 to 1000\,cm$^{-1}$) or freshly evaporated aluminum and gold mirrors were used (1000 to 10\,000\,cm$^{-1}$). The corresponding optical conductivity was calculated via  the Kramer's Kronig analysis. For the low-frequency extrapolation, in a first iteration the  Hagen Rubens relation was used, which got subsequently replaced in the fitting process; the thus determined temperature-dependence of $\sigma_\text{dc}$ follows nicely the resistivity curve (see Fig.~\ref{Parent2}). At high frequencies, the optical response was extended up to 30\,000\,cm$^{-1}$ by room temperature measurements with a Woollam spectroscopic ellipsometer; higher frequencies were extrapolated following measurements  up to 32\,eV carried out for BaFe$_2$As$_2$~\cite{Nakajima_2011_2}. For the room temperature measurements on the P substituted compounds, data were extrapolated above 10\,000\,cm$^{-1}$ with the same extrapolation as the parent compound.

\section{Results and Discussion}
\subsection{EuFe$_2$As$_2$}
\subsubsection{Overview on optical properties}
\begin{figure} [!http]
 \centering
  \includegraphics[width=1\columnwidth]{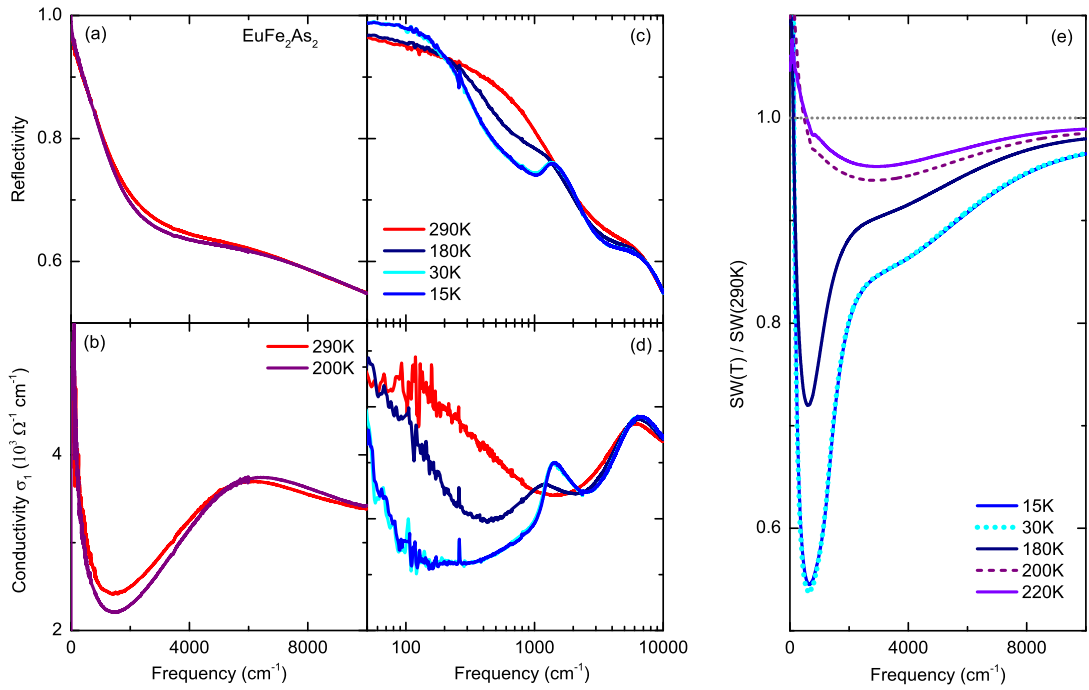}
  \caption{(a-d) Frequency-dependent in-plane reflectivity and conductivity of EuFe$_2$As$_2$ at (a-b) $T=200$\,K (purple), 290\,K (red) and (c-d)  $T=15$\,K (blue), 30\,K (light cyan), 180\,K (dark blue) and 290\,K (red),  as well as (e) spectral weight (SW) analysis for 15\,K (blue), 30\,K (light blue), 180\,K (dark blue), 200\,K (purple) and 220\,K (violet) normalized to 290\,K. With decreasing temperature, spectral weight is shifted to high energies (above $\sim 6000$\,cm$^{-1}$).  Additionally, at $T < T_\text{s,SDW}$, a gap develops in the optical spectra and spectral weight is shifted to $\sim 1400$\,cm$^{-1}$ at 15\,K; furthermore, the FeAs phonon at around 260\,cm$^{-1}$ gains considerably oscillator strength. One should note that the spectra at 15\,K and 30\,K are the same within the resolution.}
  \label{Parent1}
\end{figure}

Figs.~\ref{Parent1}a-d display the spectra for EuFe$_2$As$_2$ obtained at characteristic temperatures above and below $T_\text{s,SDW}$; while a linear scale stresses more the dynamics at higher frequencies, a logarithmic scale emphasizes the low-frequency behaviour. The overall spectral shape is characteristic for most iron pnictides. At temperatures $T > T_\text{s,SDW}$, neither reflectivity nor optical conductivity show well-separated components in the far-infrared frequency range (below $\sim 1000$\,cm$^{-1}$), leading despite a zero-energy Drude peak to a broad mid-infrared plateau that is rather untypical for a metal and often referred to as ``incoherent'' region; no clear plasma edge is visible.

At higher frequencies ($\sim 6000$\,cm$^{-1}$), a hump-like structure appears, which in principle is characteristic for interband transitions. However, varying the temperature reveals a dramatic suppression of spectral weight below $\sim 6000$\,cm$^{-1}$, which is transferred to even higher energies (see Fig.~\ref{Parent1}e). At 10\,000\,cm$^{-1}$, the total spectral weight is almost recovered\footnote{The 3\,\% decrease with lowering temperature lies within our error bar and might occur due to an underestimated narrowing of the Drude component. However, we tend to ascribe it to a real effect, as we can still resolve very small reflectivity changes at high frequencies ($< 0.5$\,\%).}. Such a high-energy spectral weight transfer is quite untypical for a conventional metal and indicates correlation effects. Intriguingly, the most significant changes appear around the energy scale associated with Hund's coupling ($J \sim 0.6 - 0.9$\,eV)~\cite{Schafgans_2012}, \textit{i.e.} the tendency of electrons to align their spins parallel~\cite{Hund}. Therefore, it was proposed~\cite{Wang_2012} that as the temperature decreases such inter-orbital coupling localizes a growing number of itinerant carriers, because their interactions with already localized electrons increase with the suppression of thermal fluctuations.

At temperatures $T < T_\text{s,SDW} = 190$\,K, a gap develops in the optical spectra, visible in a spectral weight depletion below $\sim 1000$\,cm$^{-1}$ (see Figs.~\ref{Parent1}c-e). The spectral weight is transferred to a peak in the optical conductivity around  $1400$\,cm$^{-1} \approx 10.8~ k_\text{B} T_\text{s,SDW}$, indicating the presence of an energy gap slightly below this value. The asymmetric shape of the peak is typical for a spin-density-wave gap, but could also indicate the existence of two neighbouring features (as theoretically proposed by Yin \textit{et al.}~\cite{Yin_2011}). One should further note the FeAs phonon at around 260\,cm$^{-1}$, which gains considerably oscillator strength at $T < T_\text{s,SDW}$.

Our finding of a single spin-density-wave gap in EuFe$_2$As$_2$ is consistent with our previous work~\cite{parents_2009_Wu}. We are aware that an additional smaller gap around the BCS weak coupling limit was reported by Moon \textit{et al.}~\cite{parents_2010_Moon}; this lower gap is also observed in Ba-, Sr- and Ca-122 compounds~\cite{Nakajima_2010, Hu_2008}. As we investigated a single crystal of the same batch as  Ref.~\cite{parents_2009_Wu}, one can speculate that the occurrence of the low-energy feature might strongly depend on the sample quality~\cite{Dolgov}. However, our sample has a higher residual resistivity ratio than that of Ref.~\cite{parents_2010_Moon} (see Fig.~\ref{Parent2}), indicating less impurity scattering that could smear out the gap feature.  Therefore, it is suggestive to relate this untypical behaviour to the influence of the Eu$^{2+}$ magnetism. Unfortunately, while there are predictions for the influence of the spin-density-wave on the Eu$^{2+}$ magnetism~\cite{theory_2011_Akbari}, the reverse case is -- to our best knowledge -- not yet considered theoretically. Surprisingly, the spectra at 15\,K and 30\,K are basically identical within the experimental error, meaning that Eu$^{2+}$ spin scattering seems to have a negligible direct influence on the optical properties in the infrared frequency range.

\subsubsection{Different approaches to fit the spectra}
As iron pnictides are multiband systems~\cite{multi} and reflectivity as well as conductivity typically do not show well-pronounced plasma edges, different models have been proposed to describe the data well. In order to consider which approach reflects the electrodynamics of the system best, one should note that a similar problem of a broad mid-infrared response appeared already in the case of cuprates.
There, two approaches can be identified: one-component and multi-component models~\cite{Basov_review}. In the former case, the low-frequency response is assumed to originate from itinerant carriers with a frequency-dependent scattering rate and mass (``extended Drude model''). In the multi-component approach, the conductivity spectrum is described by a free-carrier Drude term and a set of Lorentzian oscillators, accounting for example for interband transitions or impurity states. The debate which model describes better the charge dynamics in cuprates is yet not solved.
In the case of iron pnictides, the multiband character further complicates both, single- and multi-component analysis. In principle, hole and electron bands with different scattering rates should be considered. Furthermore, as five Fe $d$-bands are situated close to the Fermi energy, low-lying interband transitions are plausible (which should be subtracted before any extended Drude analysis).

Disentangling all these processes in a multi-component analysis is almost impossible and does not lead to a unique solution. Therefore, the mid-infrared response is often modeled by one rather narrow and one very broad Drude term, interpreted either in terms of electron and hole bands or coherent and incoherent contributions~\cite{Wu_2010, Barisic_2010, Nakajima_2014}. There exist also advocates of the conventional multi-component approach known from cuprates, where one or a series of Lorentzians is used to model the mid-infrared plateau. They argue that the mean free path of a very broad Drude would violate the Mott-Ioffe-Regel limit, indicating that the conductivity in such a band is no longer metallic and therefore better described by bound excitations~\cite{Tu_2010}. It turns out that in this case, enough free parameters exist in order to fit the spectrum with only one Drude -- which is again for sure not a satisfying description of a multi-band system with electron and hole bands.

\begin{figure} [!http]
 \centering
  \includegraphics[width=0.7\columnwidth]{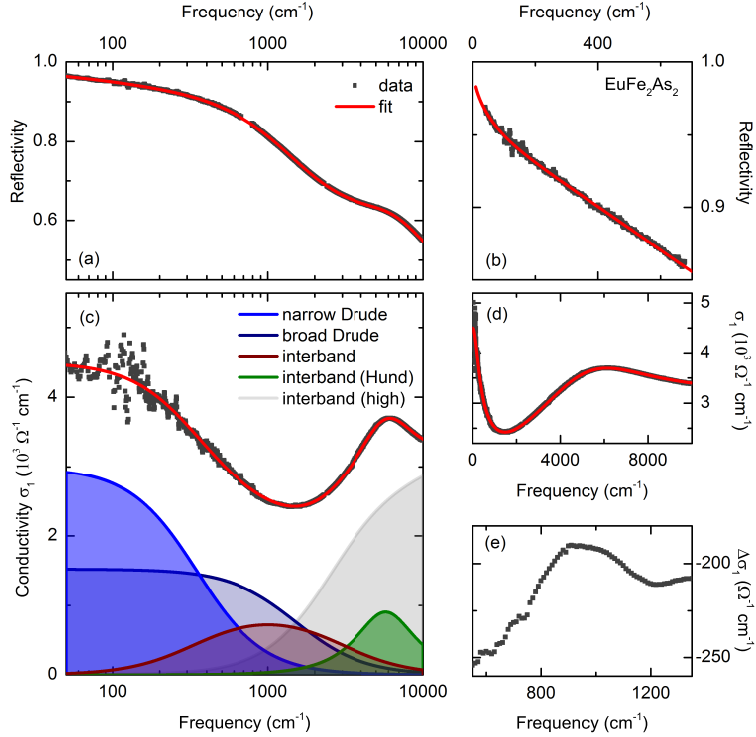}
  \caption{Frequency-dependent in-plane (a-b)~reflectivity and (c-d)~conductivity of EuFe$_2$As$_2$ at $T=$290\,K in different logarithmic and linear plot styles, as well as (e)~the conductivity difference $\Delta \sigma_1 = \sigma_1(200\text{K}) - \sigma_1(290\text{K})$. Grey squares represent data points, red lines the result of a fit with two Drude components and a Lorentzian at 1000\,cm$^{-1}$ that can be attributed to interband transitions involving Fe hole pockets; the composition of the single components is displayed in (c). The fitting approach nicely describes the spectra above $T_\text{s,SDW}$.}
  \label{Parent2a}
\end{figure}

In the present manuscript, we apply an alternative fitting procedure, motivated by recent infrared studies of Ba, Sr and Ca compounds that uncover  interband transitions around 1000 and 2300\,cm$^{-1}$ (Ref.~\cite{Marsik_2013}). Indeed, by subtracting 200\,K and 290\,K from each other, we observe similar changes around 1000\,cm$^{-1}$. We believe that such a subtraction feature can be nicely described by a Lorentzian that broadens with increasing temperature. Therefore, we model our normal state spectra by two Drude compounds as well as such a ``mid-infrared Lorentzian''. In a first approach, we only change the width of the Lorentzian with temperature; in a second variant, we have also fitted the high-temperature spectra under the condition that the broader Drude should not get too broad (and thus not violate the Mott-Ioffe-Regel limit), allowing a changing spectral weight of the mid-infrared Lorentzian. In both cases, the overall fit quality is rather good (see Fig.~\ref{Parent2a}) and the  obtained $\sigma_\text{dc}$ values perfectly follow the temperature-dependence of the resistivity (see Fig~\ref{Parent2}c).

\subsubsection{Discussion}
\begin{figure} [!http]
 \centering
  \includegraphics[width=0.5\columnwidth]{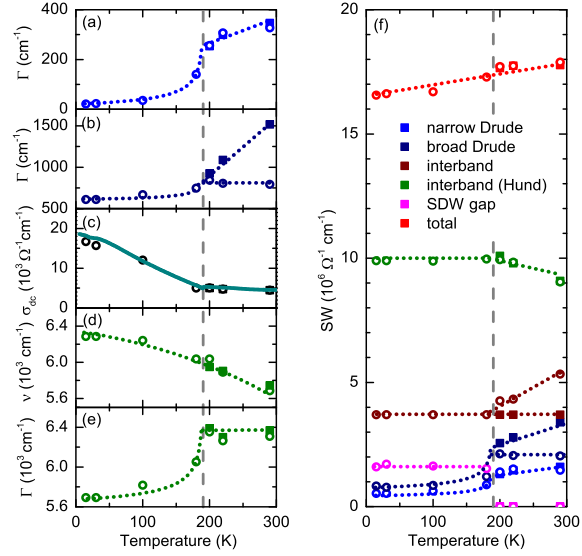}
  \caption{Temperature-dependent fit parameters used to describe the electrodynamic properties of EuFe$_2$As$_2$: (a)~Scattering rate $\Gamma \sim \tau^{-1}$ of the narrow Drude, (b)~scattering rate $\Gamma$ of the broad Drude, (c)~$\sigma_\text{dc}$ of both Drude components compared to results from resistivity measurements (data from Ref.~\cite{2014_Zapf} normalized to $\sigma_\text{dc}(300\text{K})$, cyan curve); (d)~Center frequency $\nu$ and (e)~damping rate $\Gamma$ of the feature that is usually attributed to Hund's coupling; (f) spectral weight of the temperature-dependent fitting components. The grey, vertical dashed line denotes $T_\text{s,SDW}$; other dashed lines are a guide to the eye. Closed and open symbols correspond to different fit approaches.}
  \label{Parent2}
\end{figure}

The temperature dependence of the main fit parameters is depicted in Fig.~\ref{Parent2}. The scattering rate and the spectral weight of the narrow Drude component decrease slightly with decreasing temperature, until they drop significantly at $T_\text{s,SDW}$. This description is consistent with earlier measurements on other 122 iron pnictides, indicating less scattering centres remaining in the gapped state. The broad Drude component behaves similarly; however, we can not disentangle the parameters for the broad Drude and the mid-infrared Lorentzian; a realistic description probably lies in between or rather contains several narrow temperature-dependent Lorentzians~\cite{Marsik_2013}.

Here, we want to focus on the behaviour of the bump-like feature around 6000\,cm$^{-1}$ that contains most of the spectral weight of the infrared spectra and is usually associated with Hund's coupling. With decreasing temperature, we observe an overall shift of the spectral weight from both Drude components to this feature and even higher energies (visible in a decreasing total spectral weight); its position shows a slight blueshift with decreasing temperature. Surprisingly, its width decreases rapidly at $T < T_\text{s,SDW}$ - directly reflecting the behaviour of the Drude components. This indicates a narrowing of the related bands and thus a stronger localization of the affected carriers. Furthermore, our fit indicates two different reasons for the spectral-weight transfer to very high energies, as the 6000\,cm$^{-1}$ feature itself has a conserved spectral weight below $T_\text{s,SDW}$. This observation is consistent with an interpretation in terms of Hund's coupling, because the development of a gap at the Fermi surface basically eliminates all free carriers that could get localized further. However, the possible additional spectral weight shift above 10\,000\,cm$^{-1}$ needs further investigations; from Fig.~\ref{Parent2} we speculate that - if it is real - it is related to the gapping of the broader Drude term.

\subsection{Detwinned EuFe$_2$As$_2$}
\begin{figure} [!http]
 \centering
  \includegraphics[width=1\columnwidth]{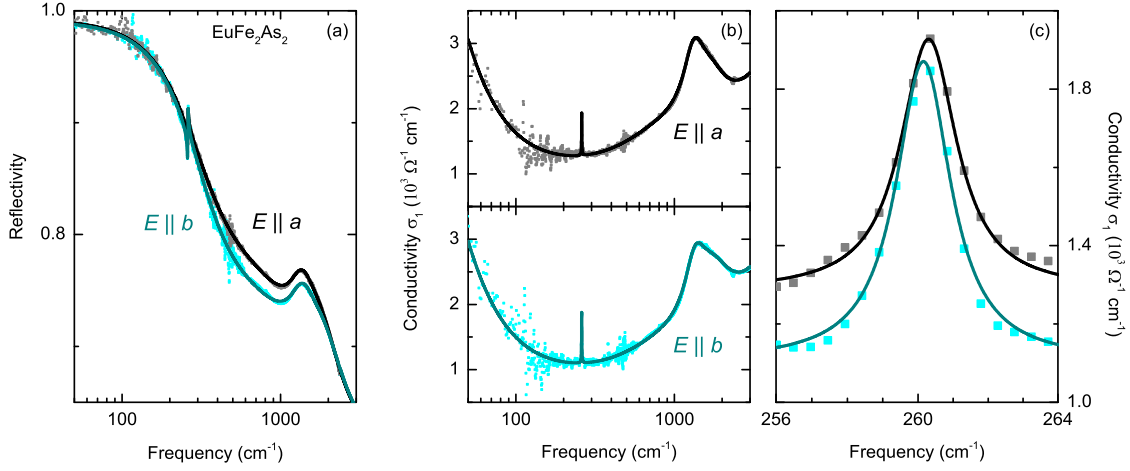}
  \caption{Frequency-dependent in-plane (a)~reflectivity and (b-c)~conductivity of EuFe$_2$As$_2$ at $T=$30\,K. Grey (light cyan) squares represent data points for $E \parallel a$ ($E \parallel b$), black (dark cyan) lines the corresponding fit with two Drude components and a Lorentzian at 1000\,cm$^{-1}$. (c)~The phonon is narrower along the $b$-direction and slightly shifted to lower frequencies.}
  \label{EuOptic_detw}
\end{figure}

We have also performed measurements on magnetically detwinned samples over a broad energy range.
As the detectors available in the very-far infrared and mid infrared energy range could not guarantee (in conjunction with the complex magneto-optical setup) a high enough signal-to-noise ratio as well as a sufficient time-stability for a complete gold evaporation run, we present here only results at 30\,K (see Fig.~\ref{EuOptic_detw}). While far-infrared data are measured completely in the magneto-optical setup, the reflectivity of the other energy ranges were obtained as followed: absolute values achieved in the conventional zero-field cryostat were normalized by the ratio of the reflectivities before and after a magnetic field of 1\,T was applied (``field treatment'', see Ref.~\cite{2014_Zapf}), both obtained in the magneto-optical setup. The very-far and mid infrared data are smoothed; higher frequencies were extrapolated with the 0\,T data, low frequencies according to the $\sigma_\text{dc}$-values.

Similar to the reports on BaFe$_2$As$_2$, we observe a strong anisotropy in the mid-infrared frequency range. Fitting the data with the same approach as introduced for the twinned samples, we find $1/\tau_b > 1/\tau_a$ and a slightly higher Drude weight for the $a$-direction, similar to Ref.~\cite{Nakajima_2012}. Furthermore, we confirm that the spin-density-wave gap related spectral weight transfer to $10.8 k_\text{B} T_\text{s,SDW}$ consists of two overlapping features, with the higher-frequency feature stronger visible in the $b$-direction. This observation is consistent with measurements and theoretical calculations for BaFe$_2$As$_2$ (Refs.~\cite{Nakajima_2011_2, Yin_2011}).

Furthermore, our observations concerning the FeAs phonon differ from the behaviour reported for BaFe$_2$As$_2$. Along the orthorhombic $a$-direction, the mode has a reduced oscillator strength, but does not disappear; neither in the here presented 0\,T data, where a detwinning fraction of approximately 75\% should appear, nor at 1\,T, where the crystal is completely detwinned~\cite{2014_Zapf}. Indeed, we are able to identify a slight blueshift of the phonon for $\sigma_a(\omega)$. This blueshift is consistent with a weak higher-energy $a$-axis phonon contribution revealed for BaFe$_2$As$_2$ by Schafgans \textit{et al.}~\cite{Schafgans_2011}. We suggest that this contribution is more pronounced in the case of EuFe$_2$As$_2$ and that the disappearance of the FeAs phonon in BaFe$_2$As$_2$ along the $a$-direction is rather accidentally and not a unique feature of iron pnictides.

\subsection{EuFe$_2$(As$_{1-x}$P$_x$)$_2$}
Recently, the Uchida group performed a systematic series of room temperature studies on electron doped Ba\-(Fe$_{1-x}$Co$_x$)$_2$\-As$_2$, hole doped Ba$_{1-x}$K$_{x}$\-Fe$_2$\-As$_2$, and isovalently substituted Ba\-Fe$_2$\-(As$_{1-x}$P$_x$)$_2$ (Refs.~\cite{Nakajima_2010_1, Nakajima_2014, Nakajima_2013}). They concluded that coherent transport is enhanced by Co and P substitution, and slightly suppressed by K doping; incoherent transport is basically not affected by Co doping, reduced by P substitution and enhanced by K doping. Therefore, it seems that chemical substitution controls the Drude spectral weight in quite different ways depending on dopant sites or type. The commonality found in these studies is that  superconductivity is suppressed when the charge dynamics become too coherent. Thus, good metallic conduction opposes superconductivity; the interaction by scattering is crucial.

\begin{figure} [!http]
 \centering
  \includegraphics[width=1\columnwidth]{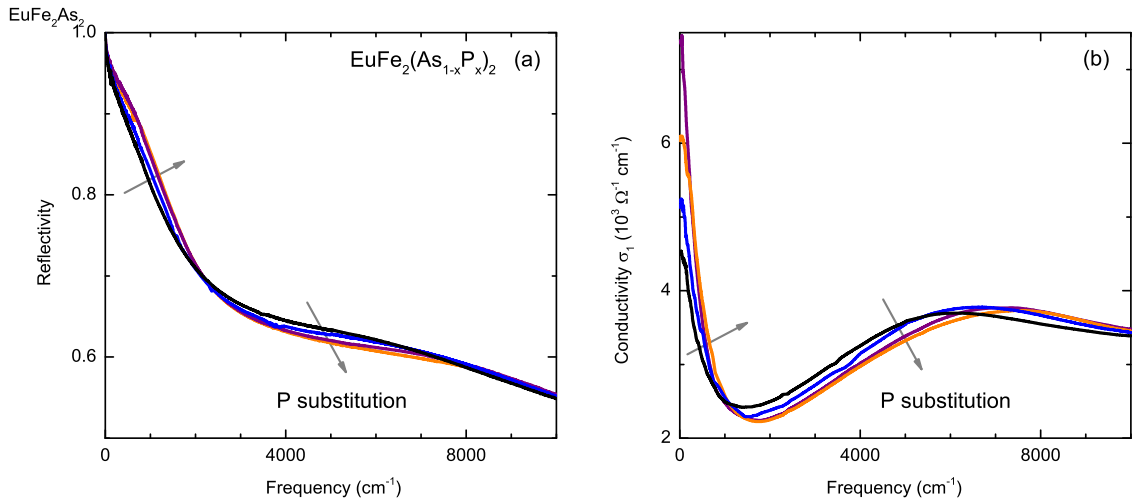}
  \caption{Frequency-dependent in-plane (a)~reflectivity and (b)~conductivity of EuFe$_2$(As$_{1-x}$P$_x$)$_2$ at $T=$290\,K for $x = 0$ (black), 0.12 (blue), 0.165 (orange) and 0.26 (purple). With increasing P substitution, reflectivity and conductivity increase at low frequencies and decrease at high frequencies, indicated by the arrows}
  \label{RSigma}
\end{figure}

\begin{figure} [!http]
 \centering
  \includegraphics[width=0.5\columnwidth]{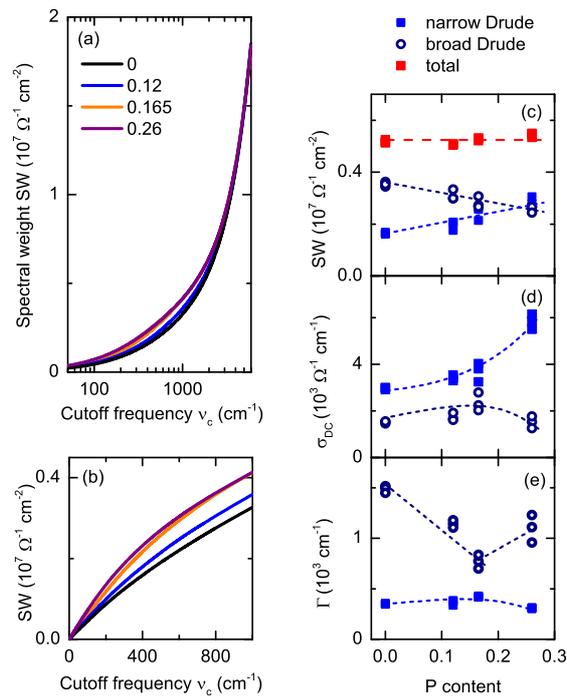}
  \caption{(a-b)~Spectral weight of EuFe$_2$(As$_{1-x}$P$_x$)$_2$ at $T=$290\,K for $x = 0$ (black), 0.12 (blue), 0.165 (orange) and 0.26 (purple), as well as the corresponding fit parameters (c)~spectral weight, (d)~$\sigma_\text{dc}$ and (e)~scattering rate $\Gamma$ of the narrow (light blue closed squares) and broad Drude (dark blue open circles). The redistribution of spectral weight takes place mainly  below 6000\,cm$^{-1}$ and corresponds to a transfer of spectral weight between the two Drude components. The scattering rate of the broad Drude components displays a non-monotonous behaviour around $x \sim 0.165$, which is probably due to a Lifshitz transition. All fits were performed three times independently, assuming constant contributions from the interband transitions around 1000\,cm$^{-1}$.}
  \label{SW}
\end{figure}

In order to elucidate the role of the spacer element between the FeAs layers, we studied the room temperature optical properties of EuFe$_2$(As$_{1-x}$P$_x$)$_2$ with $x = 0$, 0.12, 0.165 and 0.26. The corresponding reflectivity and conductivity measurements are displayed in Fig.~\ref{RSigma}. With increasing P substitution, reflectivity and conductivity increase at low frequencies and decrease at high frequencies. The redistribution of spectral weight, which takes place mainly  below 6000\,cm$^{-1}$ (see Figs.~\ref{SW}a,b), is consistent with the isovalent substitution that does not induce additional carriers. All these observations resemble those on BaFe$_2$(As$_{1-x}$P$_x$)$_2$ (Ref.~\cite{Nakajima_2013}).

For a more detailed investigation, we have also analysed our data with the modelling approach presented above, assuming constant contributions from the interband transitions around 1000\,cm$^{-1}$. The main results are summarized in Figs.~\ref{SW}c-e. This quantitative analysis further confirms that the spectral weight is redistributed from the broad to the narrow Drude. While possible scenarios for the origin of this behaviour were already extensively discussed in Ref.~\cite{Nakajima_2013}, we want to focus here more on an analysis of scattering rate and $\sigma_\text{DC}$.

At zero frequency, the conductivity changes mostly for the narrow Drude, while the contribution from the broad Drude is rather constant. The latter originates from an overall decrease in the scattering rate of the broad Drude with increasing P substitution, while the scattering rate of the narrow Drude is rather constant in our limited P substitution range. A more detailed inspection, however, reveals a non-monotonous behaviour of the scattering rate of the broad Drude around $x \sim 0.165$, just where the narrow superconducting dome in EuFe$_2$(As$_{1-x}$P$_x$)$_2$ is located. This anomaly probably originates from a Lifshitz transition at slightly higher P substitution, where one of the hole pockets disappears at the $\Gamma$-point~\cite{P_2011_Thiru}.

However, one should keep in mind that these results were obtained by assuming constant contributions from the interband transitions around 1000\,cm$^{-1}$, which is definitely oversimplified. Indeed, their spectral weight should decrease with the suppression of the hole pocket~\cite{Marsik_2013}. Therefore, it is likely that some fraction of the reduced spectral weight associated here with the broad Drude actually arises from the suppressed interband transitions. Nevertheless, as our fit model yields conserved total spectral weights of the Drude components which is consistent with isovalent substitution, this effect seems to be rather negligible.

\section{Summary}
We have compared the optical  properties of Eu- and Ba-122 iron pnictides in the normal state, based on our comprehensive infrared investigations. The overall optical response is strikingly similar, although EuFe$_2$As$_2$ possesses a significantly higher spin-density-wave transition temperature ($T_\text{SDW} \sim 190$\,K) and an additional order of the local Eu$^{2+}$ moments around $T_\text{Eu} \sim 20$\,K. In our analysis, we concentrated on electronic scattering effects. By introducing a multi-component model to describe the optical properties in the infrared spectral range, we found that a high-energy feature around 0.7\,eV is remarkably sensitive to the spin-density-wave ordering. This further confirms its direct relationship to the dynamics of itinerant carriers consistent with its interpretations in terms of Hund's rule coupling. In the case of magnetically detwinned EuFe$_{2}$As$_{2}$, we confirmed at 30\,K  a higher Drude weight and lower scattering rate along the crystallographic $a$-axis recently found for mechanically detwinned Ba-122 compounds~\cite{Nakajima_2012}. Finally, we have analysed the development of the room temperature spectra with isovalent Phosphor substitution and highlighted changes in the scattering rate of hole-like carriers induced by a Lifshitz transition.

\section{Acknowledgements}
We thank M. Daghofer and J. Fink for helpful discussions. The project was supported by the Deutsche Forschungsgemeinschaft (DFG), the National Science Foundation (NSF-DMR) and the German Academic Exchange Service (DAAD).






\bibliographystyle{elsarticle-num}
\bibliography{<your-bib-database>}







\end{document}